\documentclass[epsfig]{aa}
\usepackage{graphicx}

\def\lsimeq{{_<\atop^{\sim}}}
\def\gsimeq{{_>\atop^{\sim}}}

\begin{document}

\title{The VLA-VIRMOS Deep Field}
\subtitle{I. Radio observations probing the $\mu$Jy source population}
\author{M. Bondi\inst{1} \and P. Ciliegi\inst{2} \and G. Zamorani\inst{1,2} 
\and L. Gregorini\inst{1,3} \and G. Vettolani\inst{1} 
\and P. Parma\inst{1}
\and H. de Ruiter\inst{2} \and O. Le Fevre\inst{4} \and 
M. Arnaboldi\inst{5} \and L. Guzzo\inst{6} \and 
D. Maccagni\inst{7} \and R. Scaramella\inst{8}
\and C. Adami\inst{4} \and S. Bardelli\inst{2} \and M. Bolzonella\inst{7} 
\and D. Bottini\inst{7} \and A. Cappi\inst{2} \and S. Foucaud\inst{4} 
\and P. Franzetti\inst{7} \and B. Garilli\inst{7} \and S. Gwyn\inst{4}
\and O. Ilbert\inst{4} \and A. Iovino\inst{6} \and V. Le Brun\inst{4} 
\and B. Marano\inst{9} \and C. Marinoni\inst{4} \and H.J. McCracken\inst{9} 
\and B. Meneux\inst{4} \and A. Pollo\inst{6}  \and L. Pozzetti\inst{2} 
\and M. Radovich\inst{5} \and V. Ripepi\inst{5} \and D. Rizzo\inst{6} 
\and M. Scodeggio\inst{7} \and L. Tresse\inst{4} \and A. Zanichelli\inst{1} 
\and E. Zucca\inst{2}
}

\institute{Istituto di Radioastronomia del CNR, Via Gobetti 101, 
I-40129 Bologna, Italy 
\and
Osservatorio Astronomico di Bologna, Via Ranzani 1, 
I-40127, Bologna, Italy 
\and
Universit\`a degli Studi di Bologna, 
Dipartimento di Fisica, Viale Berti Pichat 6/2, I-40127 Bologna, Italy
\and 
Laboratoire d'Astrophysique de Marseille, Traverse  du Siphon-Les trois 
Lucs, BP8-13376 Marseille Cedex 12,  France 
\and 
Osservatorio Astronomico di Capodimonte, Via Moiariello 16, I-80127,
Napoli, Italy
\and
Osservatorio Astronomico di Brera, Via Brera 28, I-20121 Milano,
Italy
\and
Istituto di Astrofisica Spaziale e Fisica Cosmica del CNR, Via Bassini 15,
I-20133 Milano, Italy
\and
Osservatorio Astronomico di Roma, Via Osservatorio 2, I-00040,
Monteporzio Catone (ROMA), Italy
\and
Universit\`a degli Studi di Bologna, 
Dipartimento di Astronomia, Via Ranzani 1, I-40127 Bologna, Italy
}
\date{Received 16 December 2002 / Accepted 5 March 2003}
\abstract{
We have conducted a deep survey (r.m.s noise $\simeq 17\mu$Jy) with the Very
Large Array (VLA) at 1.4 GHz, with a resolution of 6 arcsec, of a 1 deg$^2$ 
region included in the VIRMOS VLT Deep
Survey. In the same field we already have multiband photometry down to
$I_{\rm AB}=25$, and spectroscopic observations will be obtained during the
VIRMOS VLT survey. The homogeneous sensitivity over the whole field has
allowed to derive a complete sample of 1054 radio sources ($5\sigma$ limit).
We give a detailed description of the data reduction and of the analysis of 
the radio observations, with particular care to the effects of clean bias and 
bandwidth smearing, and of the methods used to obtain the catalogue 
of radio sources. To estimate the effect of the resolution bias on our
observations we have modelled the effective angular-size distribution of the 
sources in our sample and we have used this distribution to simulate a
sample of radio sources. 
Finally we present the radio count distribution down to 0.08 mJy derived
from the catalogue. Our counts are in good agreement with the best fit
derived from earlier surveys, and are about 50\% higher than the counts in the
HDF. The radio count distribution clearly shows, with extremely good
statistics, the change in the slope for the sub-mJy radio sources. 
}
\maketitle
\section{Introduction}

\begin{figure*}
\includegraphics[width=10cm]{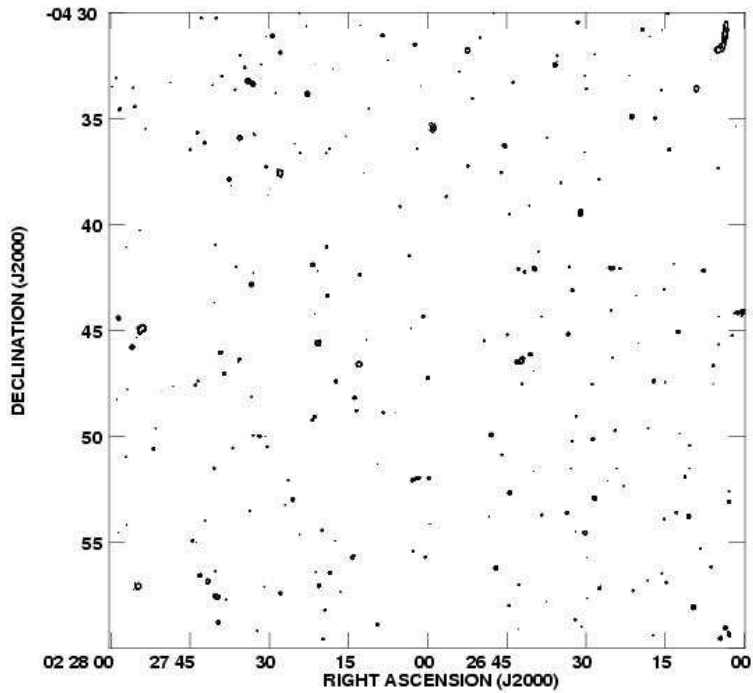}
\caption[]{South-eastern quadrant of the VLA-VDF radio image. Contours are in
units of signal to noise ratio, first contour is $5\sigma$.}
\label{field_radio1}
\end{figure*}
\begin{figure*}
\includegraphics[width=10cm]{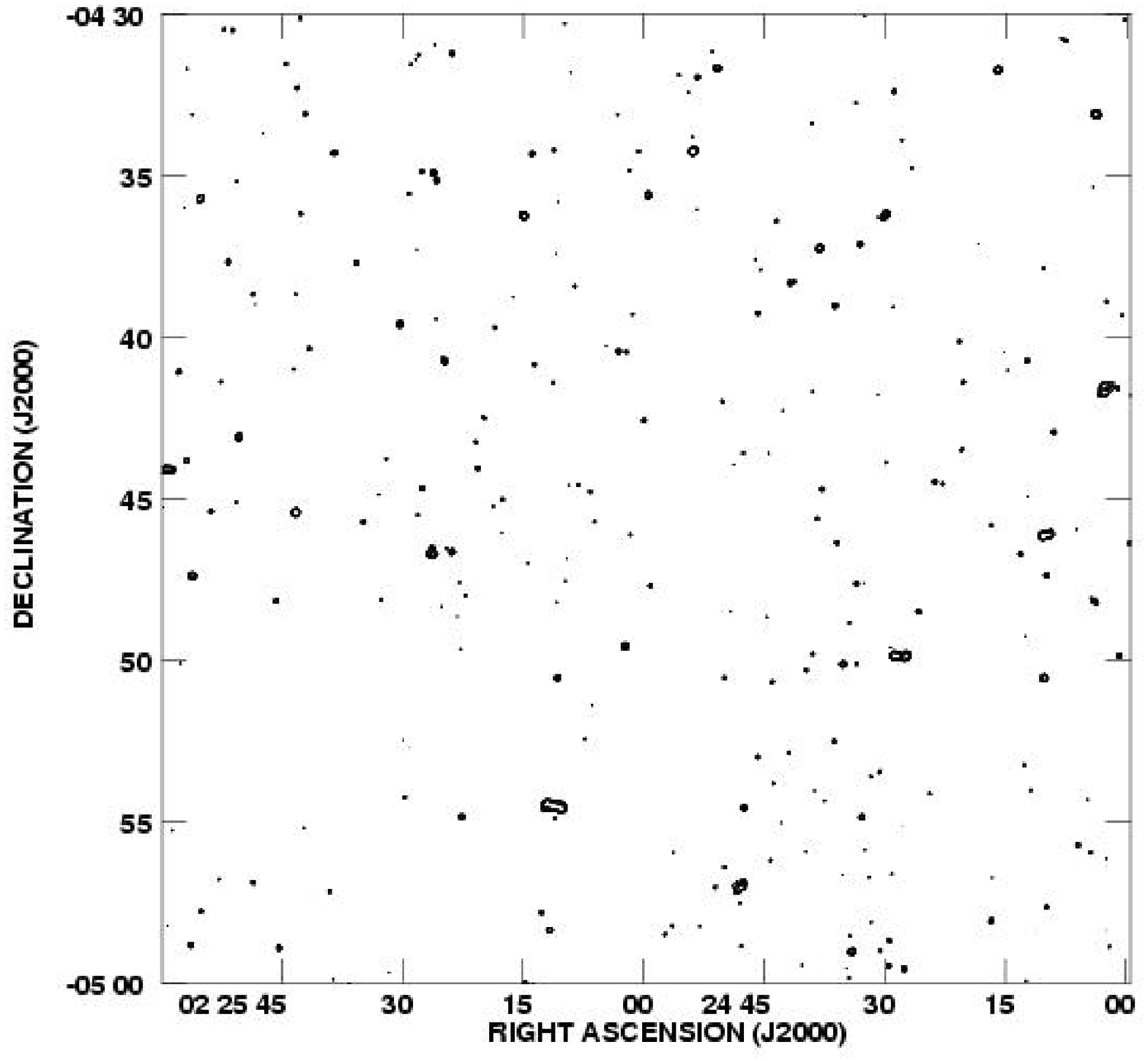}
\caption[]{South-western quadrant of the VLA-VDF radio image. Contours are in
units of signal to noise ratio, first contour is $5\sigma$.}
\label{field_radio2}
\end{figure*}
\begin{figure*}
\includegraphics[width=10cm]{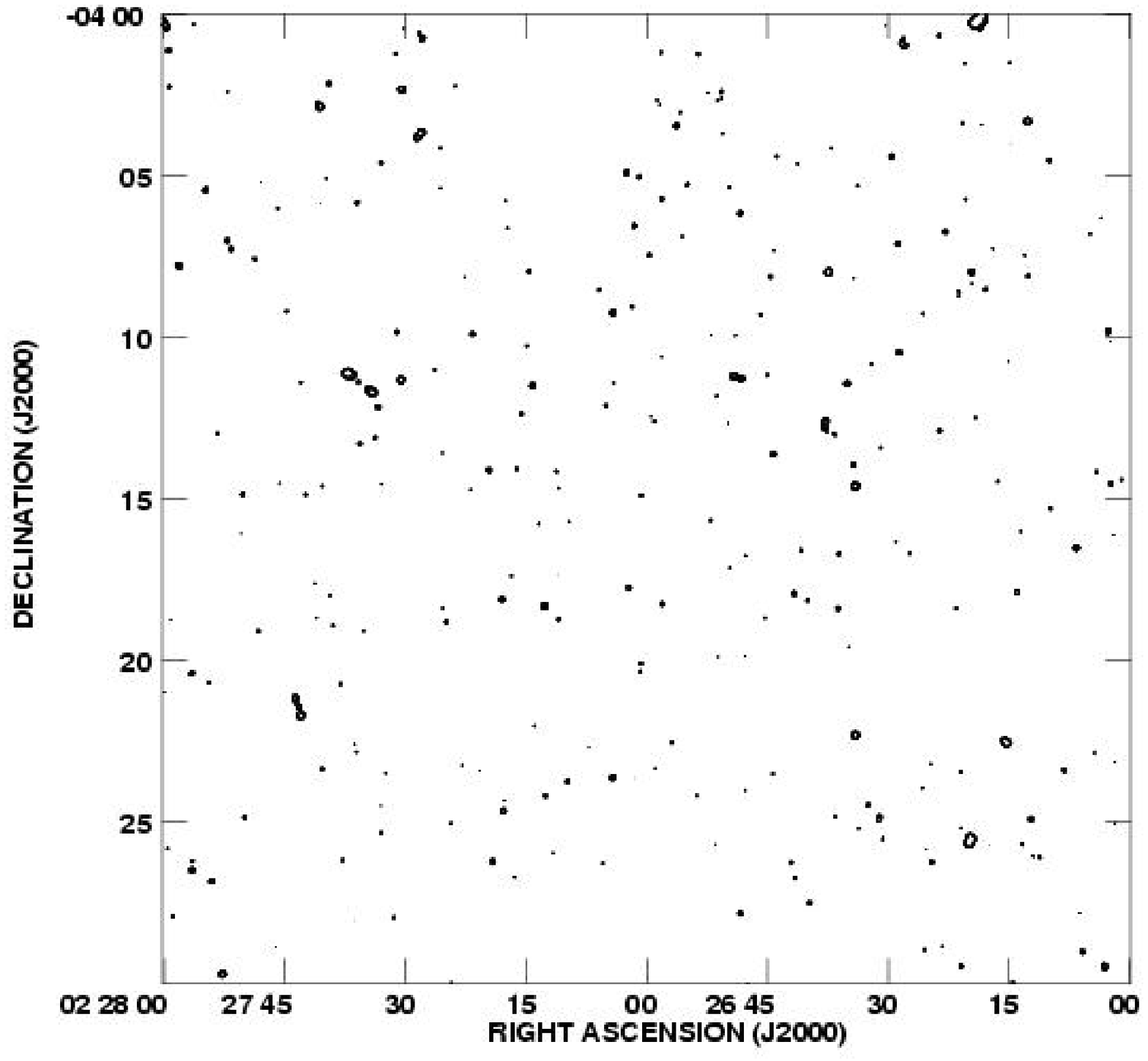}
\caption[]{North-eastern quadrant of the VLA-VDF radio image. Contours are in
units of signal to noise ratio, first contour is $5\sigma$.}
\label{field_radio3}
\end{figure*}
\begin{figure*}
\includegraphics[width=10cm]{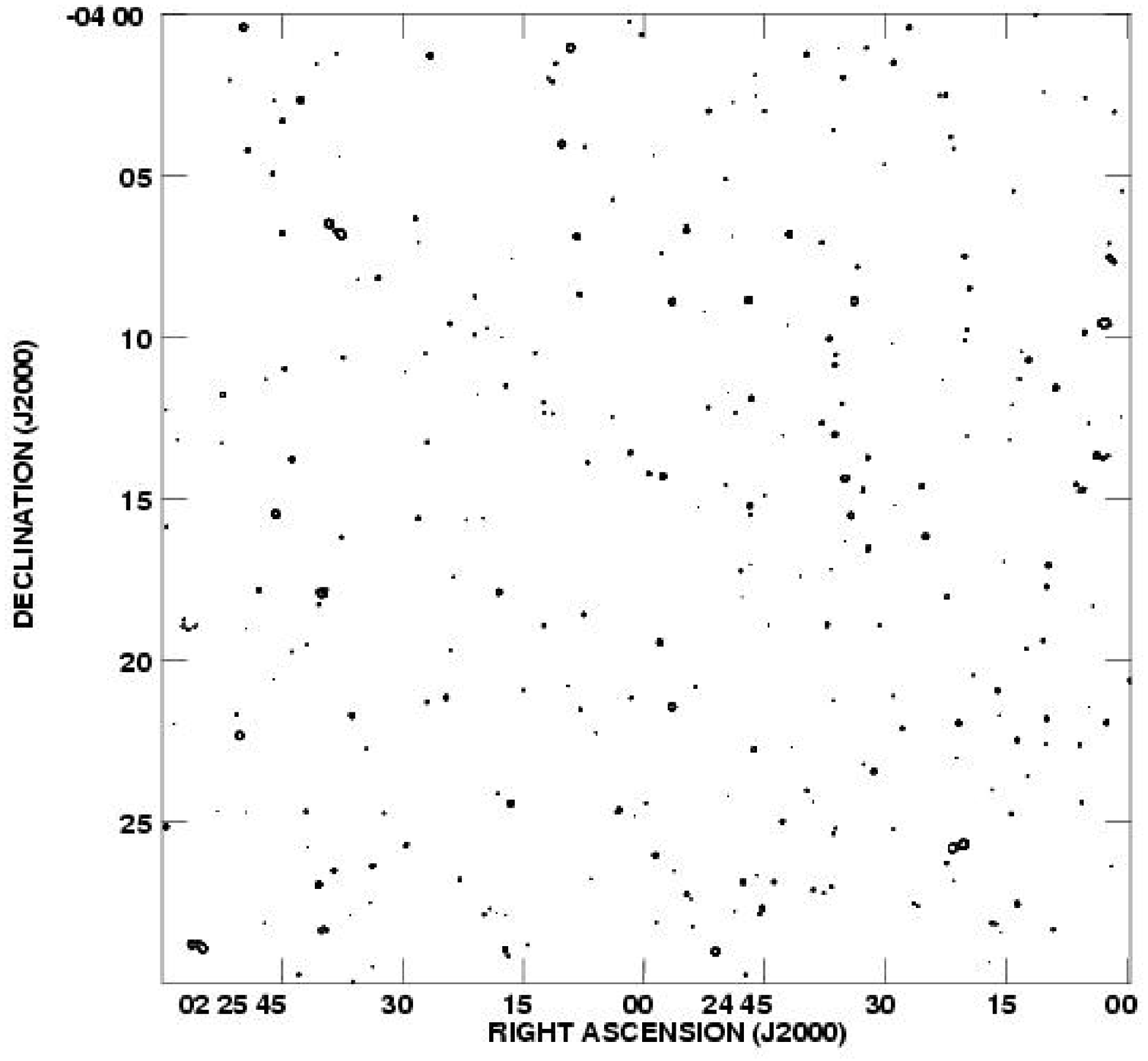}
\caption[]{North-western quadrant of the VLA-VDF radio image. Contours are in
units of signal to noise ratio, first contour is $5\sigma$.}
\label{field_radio4}
\end{figure*}

It is well established that the 1.4 GHz source counts
at sub-mJy levels reveal the
presence of a population of faint radio sources far in excess with respect
to those expected from the high luminosity radio galaxies and
quasars which dominate at higher fluxes
(\cite{Wind85};~\cite{Cond89};~\cite{Hopk98};
~\cite{Cili99};~\cite{Rich00};~\cite{Pran01a};
~\cite{Grup99b}).
Early spectroscopic studies, limited to relatively bright optical
counterparts (B$\le$21.5-22.0; \cite{Benn93}),
suggested that  most of these sub-mJy radio sources were
starburst galaxies. However it has been shown that the predominance of
starburst galaxies is dependent on the magnitude limit of the spectroscopic
follow up (\cite{Grup99a}; \cite{Pran01b}). While at bright
magnitude (B$\lsimeq 22$) most of the optical
counterparts are indeed starburst galaxies, at fainter magnitudes (B$\gsimeq
22.5$)
most of the optical counterparts appear to be early type galaxies.
This mixture of at least two different populations is consistent with what
is being found at even fainter radio fluxes, in the $\mu$Jy regime, where 
high-$z$
early type galaxies, intermediate-$z$ post starburst galaxies, and
lower-$z$ emission line galaxies are found in approximately similar proportions
(\cite{Hamm95}; \cite{Wind95}; \cite{Rich98}).

In order to fully investigate the nature and evolution of the sub-mJy
population it is absolutely necessary to couple deep radio and optical
(both imaging and spectroscopic) observations
over a reasonably large area of the sky.

The VIRMOS VLT Deep Survey (VVDS, \cite{LeFe02}) will produce
spectroscopic redshifts for about $1.5\times 10^5$ galaxies in an area of
$\sim 16$ deg$^2$
selected from an unbiased photometric sample of more than 1 million
galaxies.
We have selected a 1 deg$^2$ field from the VVDS,
centered at RA(J2000)=02:26:00 DEC(J2000)=$-04$:30:00, for deep VLA radio 
observations at 1.4 GHz (hereafter the VLA-VIRMOS Deep Field, VLA-VDF). 
This field is ideal for a radio survey 
as UBVRI photometry, complemented by K band data on a smaller region,
is already available to $I_{\rm AB} \simeq 25$ 
(\cite{LeFe01}), and spectroscopy is being obtained to $I_{\rm AB} = 24-24.5$  
with the VIMOS spectrograph at the VLT (\cite{LeFe01}). 

In this paper we present the VLA radio observations at 1.4 GHz of the 
VLA-VIRMOS Deep Field, discuss the methods used to derive the catologue
of about 1000 radio sources (down to a limit of
80 $\mu$Jy) and derive the radio  source counts. The  
identification of the
radio sources using the multi-band photometry will be discussed in a following
paper (\cite{Cili02}).
In Sect. 2 the observations and data reduction are presented.
Sect. 3 contains a detailed description of the analysis carried out
on the radio mosaic in order to quantify the effects of clean bias and
bandwidth smearing on our observations. The procedure adopted to obtain a
complete catalogue of radio sources from the radio mosaic is presented in
Sect. 4. Finally, in Sect. 5 we derive the radio counts corrected for
the resolution bias. Conclusions are given in Sect. 6.

\section{The VLA Observations}

The observations were obtained at the Very Large Array (VLA) in 
B-configuration for a total time of 56 hours 
over 9 days from November 1999 to January 2000.
This configuration was adopted as the best choice
in order to obtain a deep survey 
of a relatively large (1 square degree) area of the sky with an acceptable
resolution.

At 1.4 GHz the VLA antennas have a primary beam with a FWHM of 31 arcmin.
In order to image with uniform sensitivity a 1 square degree field it is
necessary to make multiple pointings  displaced
by about $31/\sqrt{2}\sim 22$ arcmin (\cite{Cond98}; \cite{Beck95}).
We chose to cover the surveyed area with a square grid of 9 pointings,
separated by 23 arcmin in right ascension and declination.
Such a geometry allows to reach theoretical noise variations
smaller than $10\%$ over $95\%$ of the 1 deg$^2$ field.
Each of the pointing centers was observed for a total of about 6 hours,
including the observations of the calibrators.
Every 20 minutes we interleaved the scans on the nine pointings  
with a short observation of the  source J0241$-$082 
to provide amplitude, phase, and bandpass calibration.

The observations were carried out in bandwidth synthesis mode to avoid 
substantial chromatic aberration (bandwidth smearing).
In this way it is also possible to reduce the effects of narrow-band
interferences since only the channel affected by the interferences, 
instead of the whole bandwidth, can be removed from the data.
The data were collected in spectral line mode using two intermediate frequency
(IF) bands centered at
1364.9 MHz and 1435.1 MHz. Each IF was divided in 7 channels each 3 MHz wide.
Due to limitations in the VLA correlator only circular polarization modes were
recorded.

\subsection{Calibration \& Editing}
The data were reduced and analyzed using the package AIPS developed by the
National Radio Astronomy Observatory.
The amplitude calibration was derived from daily observations of 3C~48 
assuming flux densities of 16.51 Jy at 1365 MHz and 15.87 Jy at 1435 MHz.
The AIPS tasks UVLIN and CLIP were used to flag bad visibility data
resulting from radio frequency interferences, receiver problems or correlator
failures. The task UVLIN was run after the amplitude calibration with a
threshold of 1 Jy. After UVLIN, the task CLIP was run on a channel-by-channel
basis with a threshold determined by the task UVPRM (8 times the rms of the
data).
The task UVFIX was run to improve the astrometric solution and then the
data sets for each pointing were combined and inspected for
further editing.

\subsection{Imaging, self calibration and mosaicing}

Self calibration and imaging of wide field deep observations is a 
time consuming task.
For each pointing we imaged a $2048\times 2048$ pixels area ($51\times 51$
arcminutes, 1 pixel corresponds to 1.5 arcsec) along with a number of smaller 
images (usually $32\times 32$
pixels) centered on off-axis sources that can produce confusing grating rings
in the imaged area. The possibly confusing sources have been identified with 
the RUN file generator applet available at the NVSS home page, selecting all
the sources with peak flux density greater than 1 mJy (at the NVSS resolution)
within a radius of 60 arcmin of each pointing and not included in the main 
field area.

To avoid distorsions due to the use of two dimensional FFT to approximate
the curved celestial sphere, the $2048\times 2048$ pixels area of each pointing
was not deconvolved as a single image but was split in a number of
sub-images (e.g. \cite{Perl99}).
At the end of the self-calibration deconvolution iteration scheme, the
sub-images were combined together using the AIPS task PASTE to produce
the final $2048\times 2048$ pixels image of each single pointing.
The final images have been restored with a $6\times 6$ arcseconds FWHM
gaussian beam.
We self-calibrated and cleaned the different pointings
in a way as homogeneous as
possible in order to minimize differences in the sensitivity.
Clearly, the presence or absence of relatively strong ($\simeq 10$ mJy/beam)
sources in some fields and the fluctuations in the noise produced slightly
different noise figures, with 1 $\sigma$ rms noise ranging from 14.8 to 17.9
$\mu$Jy/beam, in the nine pointings.
Finally, the 9 pointings have been combined together using the task HGEOM and 
LTESS
obtaining a linear combination weighted by the square of the beam response.
The average noise over the full 1 square degree field in the mosaic map
is 17.5 $\mu$Jy. In Fig.~\ref{field_radio1}-\ref{field_radio4} we show the 
final image of the 1 square degree VLA-VIRMOS deep field split in four 
quadrants for a clearer representation.
The noise over the 1 square degree field is homogeneous 
(see also Sect. 4.1) and the few regions devoid of sources visible in 
Fig.~\ref{field_radio1}-\ref{field_radio4} (the most notable of which is 
the area around right ascension 02:25:40 and declination -04:52:00) are real
and not artifacts produced by a much higher local noise.

\section{Analysis}

\subsection{Clean bias}
Clean bias is a recently recognized possible source of error in the flux 
density estimate derived from interferometer snapshot observations. 
The FIRST and NVSS VLA surveys, for example,
are affected by this bias and the flux densities derived from the publicly
available images have to be corrected a-posteriori.
The effect of the clean bias is that flux densities of real sources are
sistematically underestimated  because the CLEANing algorithm subtracts flux
from real sources and redistributes it on top of noise peaks or sidelobes.
This effect is dependent on the image noise and synthesized beam sidelobe
levels and mostly independent from the source flux density (\cite{Beck95};
\cite{Cond98}).

The VLA-VDF observations are very long compared to the snapshots of the VLA
surveys, and consequently the synthesized beam has much lower sidelobes.
For instance, the NVSS observations have a synthesized beam with sidelobes
reaching about $25\%$ of
the main lobe, while the sidelobes in the synthesized beam of the VLA-VDF
observations reach only $1.6\%$ of the main lobe.
A good rule of thumb to avoid clean bias is to minimize the clean area and
not to clean the image down to the theoretical noise. The first prescription
is hard to follow as our goal is to image a 1 square degree field. On the
other hand, we used a rather conservative approach halting the clean process
when the clean residuals were between 2 and 5 times the
theoretical r.m.s. noise. While we can expect
that the clean bias in  our observations is much lower than that
affecting snapshot observations, we can not rule it out completely
a-priori.
In order to assess the impact of the clean bias on the VLA-VDF observations we
have used two different methods.

The first method is a step-by-step clean.
We can expect that beyond a threshold value in the
number of iterations, the clean bias begins to be important and the
flux density of the components in these maps becomes sistematically lower than
that of the corresponding components in the images  with less iterations.
About $4\times 10^4$ clean iterations are the minimum number necessary to
effectively approach the expected noise
for the $2048\times 2048$ pixel images.
We have then chosen one of the nine fields and produced
different images with an increasing number of clean iterations.
We have produced images  with $4\times 10^4$, $5\times 10^4$,
$7.5\times 10^4$ and $5\times 10^5$
iterations and on each of them we have identified radio components
down to the $3\sigma$ limit. For each component we have extracted the peak 
flux and computed the difference of fluxes obtained between maps with  
different iteration limits.
The median of these differences is listed in 
Table~\ref{clean_bias_tab}.  
\begin{table}
\caption{Clean bias simulation}
\begin{tabular}{ccc}
\hline
\label{clean_bias_tab}
N$_{\rm iter}$(1)$-$N$_{\rm iter}$(2) & median $\Delta S$ \\
\hline
 $40000-50000$        & ~0.5 $\mu$Jy/beam\\
 $50000-75000$        & ~1.0 $\mu$Jy/beam\\
 $75000-500000$       & 14.4 $\mu$Jy/beam\\
\hline
\end{tabular}
\end{table}
As can be seen from Table~\ref{clean_bias_tab},
increasing the number of iterations to $5\times 10^4$ or even $7.5\times 10^4$
produces a systematic decrease in the peak flux densities of
about 1.0 $\mu$Jy/beam or less. We have to push the clean to a far higher
number of iterations to start seeing a significant effect due to the clean 
bias. Using $5\times 10^5$ iterations we find a
decrease in the peak flux density of the sources with a median of
14.4 $\mu$Jy/beam. We have then chosen a limit of about $5\times 10^4$ clean 
iterations for each of the nine fields.
This means that any clean bias affecting
our brightness measurements should be less than 1 $\mu$Jy/beam and then
practically negligible.

The second method used to verify the absence of a significant effect 
produced by the clean bias was
to insert artificial sources with known flux in the {\it uv}-data set of a
chosen field.  In particular, we have modified the {\it uv}-data set of a 
randomly
chosen field adding 25 point sources with 0.5 mJy flux.
We have then cleaned the field to the same depth used for the
original one and compared the fluxes of the artificial sources on the map with
their true fluxes. We have repeated this operation four times obtaining a set
of 100 artificial sources at different positions. The mean of the peak flux 
density distribution
of these 100 sources is $0.503\pm 0.003$ mJy/beam.
We can conclude that 
both methods
confirm that the flux density derived from our images are not
affected by the clean bias.

\subsection{Bandwidth Smearing}

\begin{figure}
\includegraphics[width=8cm]{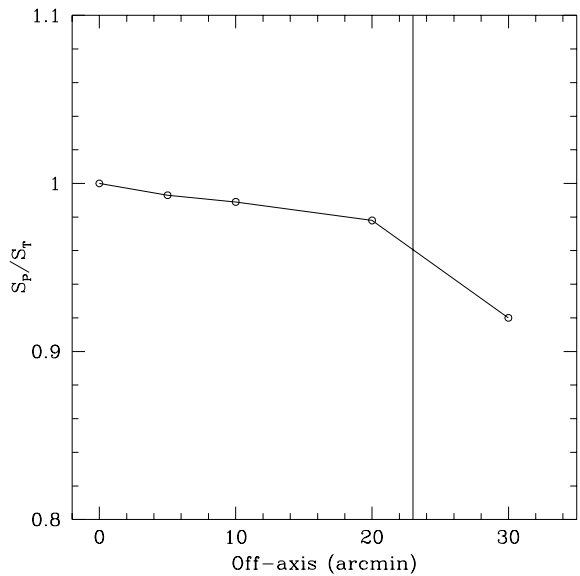}
\caption[]{Ratio between $S_{\rm P}$ and $S_{\rm T}$ for 3C 84 as a
function of off-axis angular distance  as a test of
bandwidth smearing for the VLA-VDF observations. The vertical line
indicates the distance between adjacent pointing centers.} 
\label{smearing}
\end{figure}

Imaging sources at large distances from the phase center can result in
radial smearing reducing the peak flux density of a source while conserving
its integrated flux density. This effect is known as bandwith smearing (or
chromatic aberration) and affects all the synthesis observations made with
a finite bandwidth. The image smearing is proportional to the bandwidth and to
the distance of the source from the phase center.
In order to image a 1 square degree field we have to minimize the effect of
bandwith smearing and for this reason we observed in spectral line mode.
Nonetheless, some amount of smearing can still
be present in our images.
To check this effect on our observations we have
observed the radio source 3C84 at different position offsets (0, 5, 10, 20,
30 arcmin) in two orthogonal directions. The images of 3C84 at different
offsets have been fitted with a two dimensional gaussian to derive the peak
and integrated flux density. The mean ratio between these two quantities
is shown in Fig.~\ref{smearing}. At the distance between the pointing
centers (23 arcmin) $S_{\rm P}/S_{\rm T}\sim 0.96$. Since the nine radio fields
are combined together weighted by the square of the beam response,
we can conclude that the effect of the bandwidth smearing is negligible.

\section{From the radio mosaic to the catalogue}

\subsection{The noise map} 

\begin{figure}
\begin{center}
\includegraphics[width=8cm]{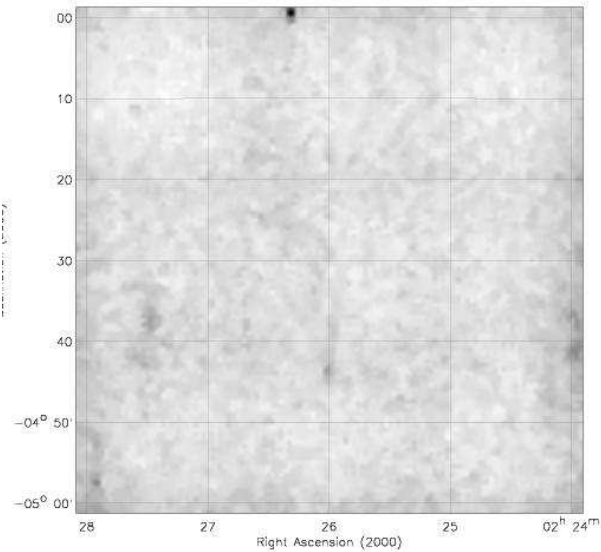}
\caption[]{A grey scale of the noise map (1 $\times$ 1 deg$^2$)
obtained with SExtractor. Darker regions mean higher noise.} 
\label{noise_map}
\end{center}
\end{figure}

\begin{figure}
\includegraphics[width=8cm]{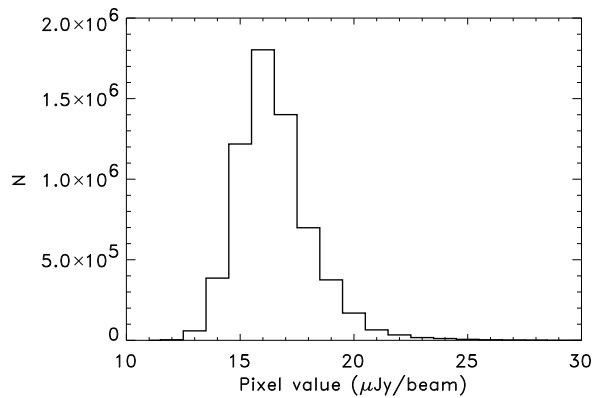}
\caption[]{Distribution of the pixel values of the noise map 
obtained with SExtractor} 
\label{histo_noise_map}
\end{figure}

Having verified that clean bias and radial smearing do not significantly 
affect the
determination of the flux densities in the VLA-VDF, we
started to work on the extraction of a catalogue from the mosaic image.
In order to select a sample above a given threshold, defined in terms
of local signal to noise ratio, we performed a detailed analysis of the
spatial root mean square (rms) noise distribution over the entire mosaic image
using the software package SExtractor (\cite{BA96}).
To construct the noise map,
SExtractor makes a first pass through the pixel data, computing an
estimator for the local background in each mesh of a grid that covers the
whole frame (see \cite{BA96} for more details).
The choice of the mesh size is very important. When it is too small,
the background estimation is affected by the presence of real sources.
When it is too large, it cannot reproduce the small scale
variations of the background. For our radio mosaic we adopted a mesh size of
20 pixels, corresponding to 30 arcsec. A grey scale of the noise map
obtained with SExtractor is shown in Fig.~\ref{noise_map}, while
Fig.~\ref{histo_noise_map} shows the histogram of its pixel values.
Due to the uniform noise over the whole field the pattern of the 9
pointings used for our observations can be barely seen in 
Fig.~\ref{noise_map}. The areas of higher noise
(black pixels) are due to the presence of
relatively strong radio sources (10-20 mJy/beam) in the map.
The pixel values distribution has
a peak at 16 $\mu$Jy/beam, well in agreement with the noise values found in 
Sect. 3.2. 

\begin{figure}
\includegraphics[width=8cm]{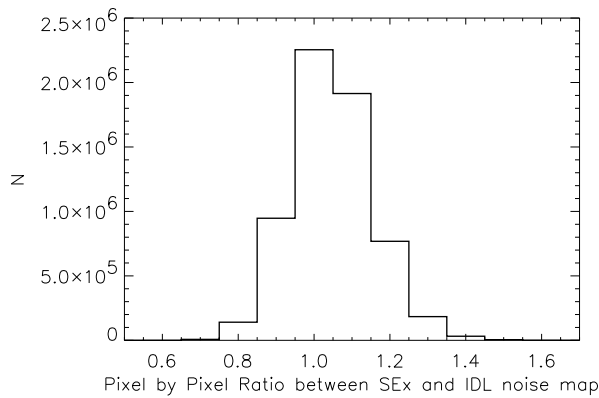}
\caption[]{Distribution of  pixel by pixel ratio  between the noise map 
obtained with SExtractor and the noise map obtained with our IDL 
code} 
\label{histo_ratio_noise_map}
\end{figure}

Since SExtractor was developed for the analysis of optical data
and this is the first attempt in using it on a radio map, we tested
the reliability of its output by constructing a noise map with
a completely independent software written in IDL language.
Briefly, starting from the residual map obtained from the AIPS task SAD
where  all the sources with peak flux greater than 60 $\mu$Jy
($\sim 3.5 \sigma$)
have been subtracted, we have first removed the most anomalous residual pixels
(including extended sources not found or rejected by SAD) substituting
these values with the average rms obtained from the residual map.
Then, we applied a local sigma clipping, substituting all the pixel values
greater than 3 times the local sigma with a random value extracted
from a Gaussian with mean and sigma equal to the local values.
Finally, the noise map  has been obtained substituting
each pixel value with the standard deviation values calculated in a local box
around each pixel (we used a box of 20$\times$20 pixels).
In Fig.~\ref{histo_ratio_noise_map} we show the distribution of the ratio
(pixel by pixel) between  the noise map
obtained with SExtractor and that obtained with our IDL code.
A detailed analysis of the two maps shows that they
are in very good agreement with each other, with differences of pixel values
which are smaller than $25\%$ over about $96\%$ per cent of the area. 

On the basis of this comparison we concluded that the noise map obtained 
with SExtractor is indeed reliable and we used it for the extraction of a 
catalogue. 

\subsection{The source detections} 

The area from which we have extracted the complete sample of sources 
is 1$\times$1 deg$^2$ centered at RA=02:26:00 DEC=$-04$:30:00 (J2000). 
Within this region we extracted all the radio components with a peak flux 
S$_{\rm P}>$60 $\mu$Jy ($\sim 3.5 \sigma$) 
using the AIPS task SAD (Search And Destroy), which attempts 
to find all the components whose peaks are brighter than a given level. 
For each selected component, the peak and total fluxes, 
the position and the size are 
estimated using a Gaussian fit. However, for faint components the Gaussian 
fit may be unreliable and a better estimate of the peak flux 
S$_{\rm P}$ (used for the selection) and of the component 
position is obtained with 
a simple interpolation of the peak values around the fitted position. 
Therefore, starting from the SAD positions, we derived the  peak flux 
S$_{\rm P}$ and the position of all the components using a second-degree 
interpolation with task MAXFIT. 
Only the components for which the ratio between the MAXFIT peak flux density 
and the local noise (derived from the noise map described in the previous 
section) was greater or equal to 5 have been included in the 
sample.  A total of 1084 components have been 
selected with this procedure. 

Some of these components clearly belong to a single radio source (e.g. the
lobes of a few FR~II radio sources, or components in very extended sources), 
but for most of them it is necessary to
derive a criterion as general as possible to discriminate between different
components of the same radio source or truly different radio sources.
For this reason, the components with distance less than 18 arcseconds
(3 times the beam size) have been
selected as possible doubles and have been visually checked one-by one using
also preliminary deep optical images. Based on the comparison of the radio
and optical fields,  we have assumed that
when two components have a distance smaller than 18 arcsec, a flux ratio
smaller
than 3, and both components have peak brightness larger than 0.4 mJy/beam
they belong to the same radio source, otherwise are considered as separate
radio sources. With this choice, considering the number of sources with flux 
greater than 0.4 mJy we can expect 3 spurious couples of radio components 
in the 1 deg$^2$ field, compared with the 40 observed. On the other hand,
within 18 arcsec
we expect 50 and find 51 couples of components with both fluxes less than
0.4 mJy.
               
The final catalogue lists 1054 radio sources, 19 of which are
considered as multiple, i.e. fitted with at least two separate
components, and it will be available on the web at
http://virmos.bo.astro.it/radio/catalogue.html.
A sample page of the
catalogue is shown in Table~\ref{sample_cat}.

\begin{table*}
\caption[]{Radio catalogue: sample page}
\vskip -3cm
\includegraphics[]{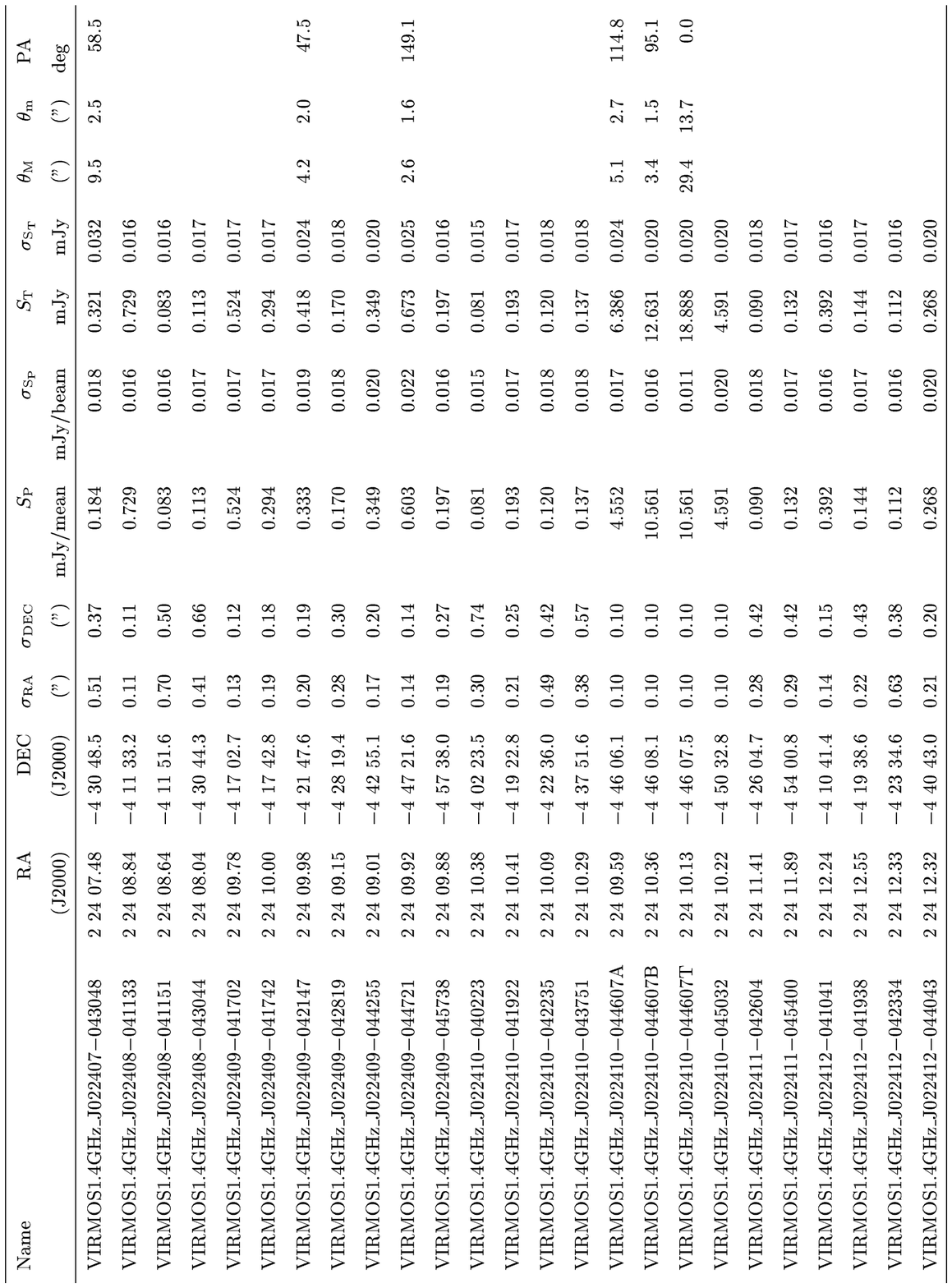}
\label{sample_cat}
\end{table*}

\begin{figure}
\includegraphics[width=8cm]{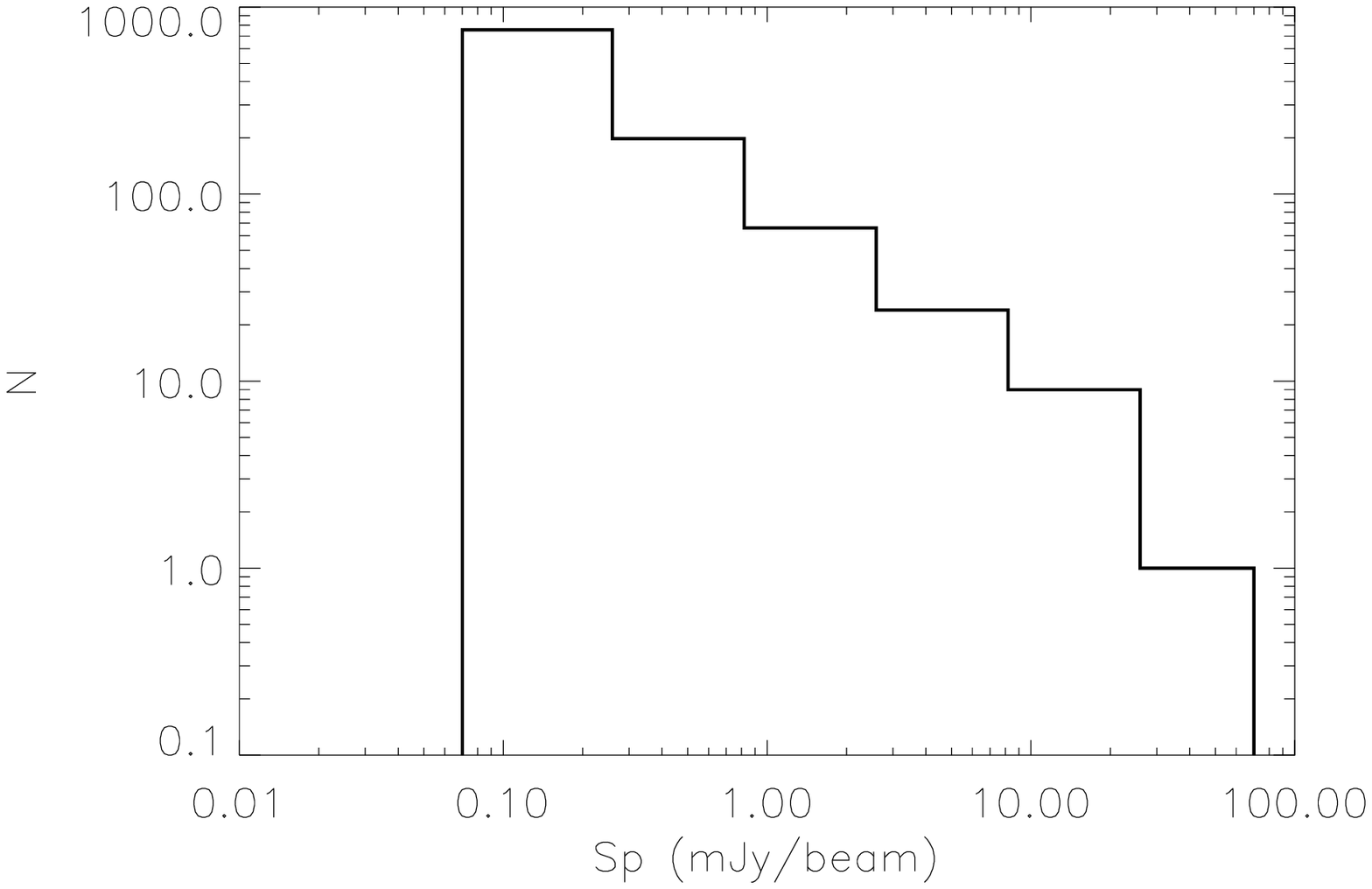}
\caption[]{Peak flux density distribution for all the 1054 
VIRMOS radio sources}
\label{Speak_distribution}
\end{figure}

For each source we list
 the source name, position in RA and DEC with errors, peak flux and
 total flux density with errors,
major and minor axis and position angle.
For the unresolved sources the total flux density is equal to the peak
brightness and the angular size is undetermined.
For each of the 19 sources fitted with multiple components we list in the
catalogue an entry for each of the components, identified with a trailing 
letter (A, B, C, {\ldots} ) in the source name, and an entry for the whole
source, identified with a trailing T in the source name. 
In these cases the total flux was calculated 
using the task TVSTAT, which allows the integration of the map values over
irregular areas, and the sizes are the largest angular sizes. 
The peak flux density distribution
of the 1054 radio sources is shown in Fig.~\ref{Speak_distribution}.

\subsection{Resolved and Unresolved Sources} 

Since the ratio between total and peak fluxes is a direct measure 
of the extension of a radio source, we used it to discriminate between 
resolved or extended sources ($i.e.$ larger than the beam) and 
unresolved sources. 

\begin{figure}
\includegraphics[width=8cm]{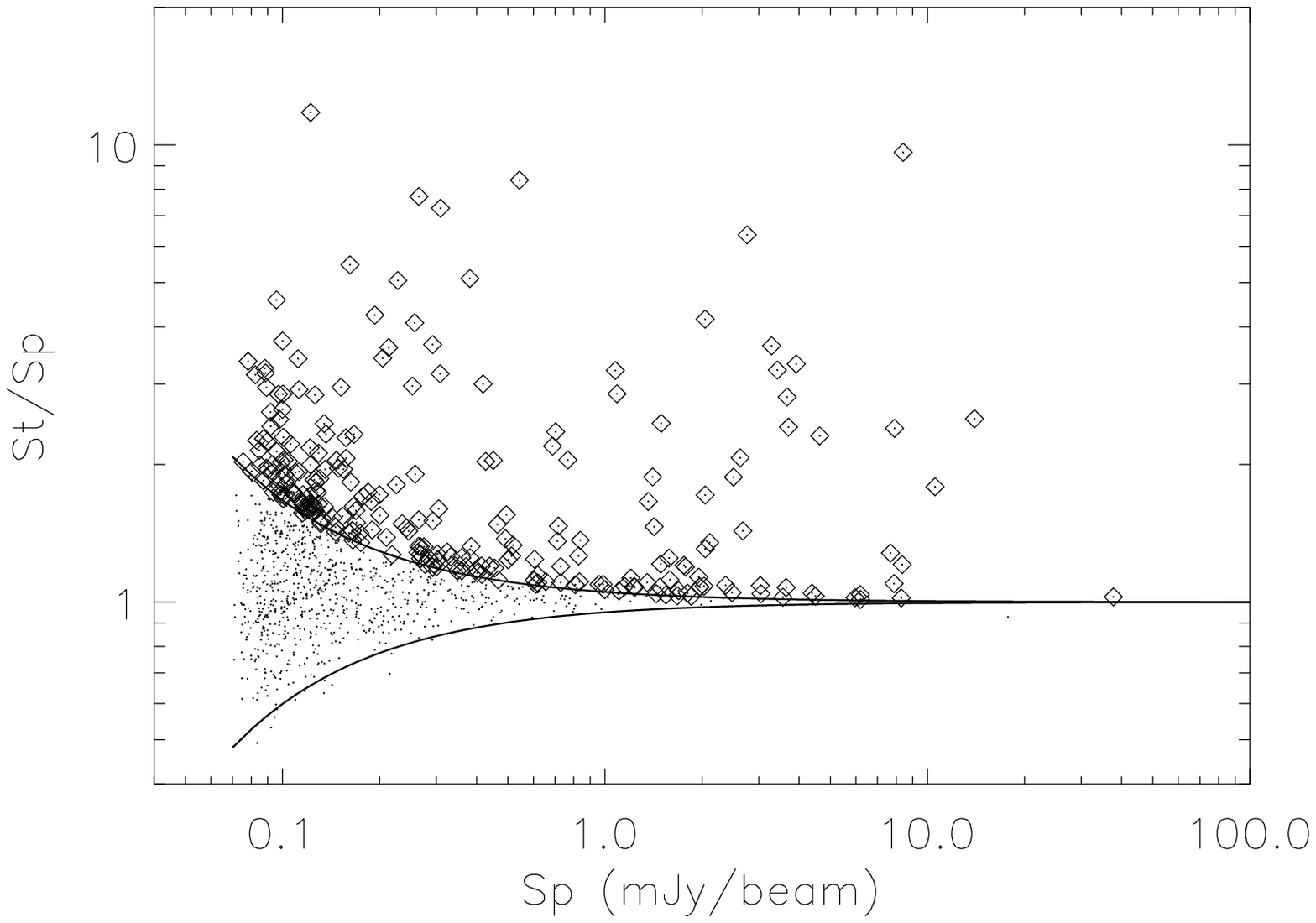}
\caption[]{Ratio of the total flux S$_{\rm T}$ to the peak flux 
S$_{\rm P}$ as a function of S$_{\rm P}$. The solid lines
show the upper and lower envelopes of the flux ratio distribution 
containing the sources considered unresolved (see text). Open simbols 
show the sources considered extended.} 
\label{ratio_fit_allflux}
\end{figure}

In Fig.~\ref{ratio_fit_allflux} we plot the ratio  between
the total S$_{\rm T}$ and the peak S$_{\rm P}$ flux density as a function of 
the peak flux
density for all the radio sources in the catalogue. To select the resolved
sources, we have determined the lower envelope of the flux ratio
distribution of Fig.~\ref{ratio_fit_allflux} and, assuming that values of
S$_{\rm T}$/S$_{\rm P}$ smaller than 1 are purely due to statistical errors, 
we have mirrored it above the  S$_{\rm T}$/ S$_{\rm P}$=1 value (upper 
envelope in Fig.~\ref{ratio_fit_allflux}). We have considered extended the 
254 sources laying above the upper envelope, that can be characterized by
the equation 

\begin{equation}
\frac{S_{\rm T}} {S_{\rm P}} = 0.95^{-(1/{\rm S_P})}
\end{equation}

\subsection{Errors in the source parameters}

The formal relative errors determined by a Gaussian fit are generally smaller 
than the true uncertainties of the source parameters. Gaussian random noise 
often dominates the errors in the data (\cite{Cond97}). Thus, we used the
Condon (1997) error propagation equations to estimate the true errors on
fluxes and positions:

\begin{equation}
\frac{\sigma^2_{\rm S_P}}{S_{\rm P}^2}=\frac{\sigma^2_{\rm S_T}}
{S_{\rm T}^2}=\frac{2}{\rho^2}
\end{equation}

 where S$_{\rm P}$ and $S_{\rm T}$ are the peak and the total fluxes, 
 and $\rho$ is the signal--to--noise ratio, given by

\begin{equation}
\rho^2 = \frac{\theta_{\rm M} \theta_{\rm m}}{4\theta_{\rm N}^2} 
\left[ 1+ \left(
\frac{\theta_{\rm N}}{\theta_{\rm M}} \right)^2 \right]^{\alpha_{\rm M}} 
\left[ 1+ \left(
\frac{\theta_{\rm N}}{\theta_{\rm m}} \right)^2 \right]^{\alpha_{\rm m}} 
\frac{S_{\rm P}^2}
{\sigma_{\rm map}^2}
\end{equation}

where $\theta_{\rm M}$ and $\theta_{\rm m}$ are the
fitted FWHMs of the major and minor axes,
$\sigma_{map}$ is the noise variance of the image and $\theta_{\rm N}$ is
the FWHM of the Gaussian correlation length of the image noise ($\simeq$FWHM
of the synthesized beam).

The exponents are  $\alpha_{\rm M} = \alpha_{\rm m} = 3/2$.

The projection of the major and minor axis errors onto the right ascension and
declination axes produces the total rms position errors given by
\cite{Cond98}:

\begin{equation}
\sigma^2_{\alpha} = \varepsilon^2_{\alpha} + \sigma^2_{\rm x_0} \sin^2(PA) +
\sigma^2_{\rm y_0} \cos^2(PA) \label{eq:ra_rms}
\end{equation}
\begin{equation}
\sigma^2_{\delta} = \varepsilon^2_{\delta} + \sigma^2_{\rm x_0} \cos^2(PA) +
\sigma^2_{\rm y_0} \sin^2(PA) \label{eq:dec_rms}
\end{equation}

where PA is the position angle of the major axis,
($\varepsilon_{\alpha}, \varepsilon_{\delta}$) are the `calibration''
errors, while $\sigma_{\rm x_0}$ and $\sigma_{\rm y_0}$ are 
$\theta_{\rm M}^2/(4ln2)\rho^2$
and $\theta_{\rm m}^2/(4ln2)\rho^2$ respectively.

\begin{figure}
\includegraphics[width=8cm]{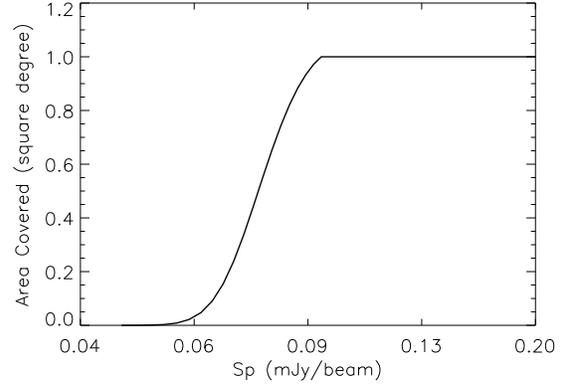}
\caption[] {Visibility area of the VIRMOS radio survey. It represents the
area over which a source with a peak flux density S$_{\rm P}$ can be detected.}

\label{area_limit}
\end{figure}

Calibration terms are in general estimated from comparison with 
external data of better accuracy than the one tested, using
sources strong enough that the noise terms in equation~\ref{eq:ra_rms}
and~\ref{eq:dec_rms} are much smaller than the calibration terms.
Unfortunately there
are no such data available in the region covered by the VLA-VDF survey.
In fact, the only other radio data available in this region are the
NVSS radio data, with a synthesized beam of 45$^{\prime\prime}$,
about a factor 8 greater than the synthesized beam of the VLA-VDF survey.
To estimate the calibration terms $\varepsilon_{\alpha}$ and
$\varepsilon_{\delta}$
we have selected all the point sources with S$_{\rm P} \ge 0.3$ mJy/beam
from the final mosaic (105 objects) and compared their positions with
those found on the single images.
The mean values and standard deviations found from this
comparison are $<\Delta$RA$>=0.007\pm 0.076$ arcsec
and $<\Delta$DEC$>=-0.039\pm 0.115$ arcsec. These values are consistent with
no systematic offset in right ascension and declination
and a standard deviation of about 0.1 arcsec. In calculating the errors
affecting the radio position of the sources in the catalogue we have
assumed $\varepsilon_{\alpha}\simeq \varepsilon_{\delta}\simeq 0.1$ arcsec.
\begin{figure}
\includegraphics[width=8cm]{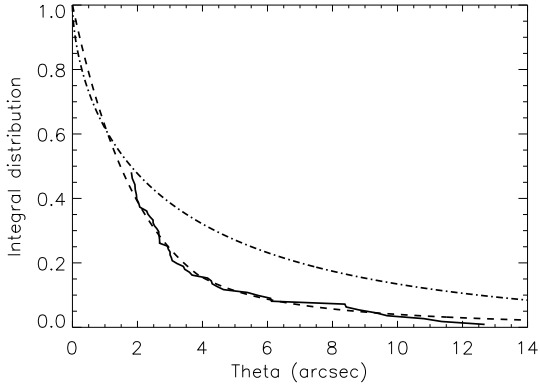}
\caption[]{Integral angular size distribution (solid line) for
the sources in the
VLA-VDF survey with 0.4$\leq$S$<$1.0 mJy. The dashed line shows
the best fit obtained for this distribution
(h($\theta$)=1/(1.6$^{\theta}$) for $\theta\leq4^{\prime \prime}$
and h($\theta$)= ($\theta^{-1.3} - 0.01)$ for $\theta > 4^{\prime \prime}$),
while the dot-dashed line shows   the integral angular
distribution reported by Windhorst et al. (1990)
and assumed by several authors as the true angular size distribution
of radio sources.
}
\label{theta_int_dist}
\end{figure}

\begin{table*}
 \centering
   \caption{The 1.4 GHz Radio Source Counts for the VIRMOS survey}
     \label{counts_tab}
     
     \begin{tabular}{ccccccc}
     
     \hline

$S$  & $<S>$ & $N_S$ & $dN/dS$             & $nS^{2.5}$          & $C$  &  $N(>S)$ \\
(mJy) & (mJy) &       & sr$^{-1}$ Jy$^{-1}$ & sr$^{-1}$ Jy$^{1.5}$ &      &deg$^{-2}$ \\
\hline
 ~0.08 -- ~~0.12 &   0.10 & 346 & 3.07$\times10^{10}$ & ~2.91$\pm$~0.16   &  1.29&1200  \\
 ~0.12 -- ~~0.18 &   0.15 & 205 & 1.12$\times10^{10}$ & ~2.94$\pm$~0.21   &  1.25&~718  \\
 ~0.18 -- ~~0.27 &   0.22 & 167 & 6.09$\times10^9$    & ~4.40$\pm$~0.34   &  1.00&~462  \\
 ~0.27 -- ~~0.41 &   0.33 &  99 & 2.41$\times10^9$    & ~4.79$\pm$~0.48   &  1.00&~295  \\
 ~0.41 -- ~~0.61 &   0.50 &  58 & 9.40$\times10^8$    & ~5.15$\pm$~0.68   &  1.00&~196  \\
 ~0.61 -- ~~0.91 &   0.74 &  42 & 4.54$\times10^8$    & ~6.85$\pm$~1.06   &  1.00&~138  \\
 ~0.91 -- ~~1.37 &   1.12 &  25 & 1.80$\times10^8$    & ~7.50$\pm$~1.50   &  1.00&~~96  \\
 ~1.37 -- ~~2.05 &   1.67 &  18 & 8.65$\times10^7$    & ~9.91$\pm$~2.34   &  1.00&~~71  \\
 ~2.05 -- ~~3.08 &   2.51 &  17 & 5.44$\times10^7$    & 17.20$\pm$~4.17   &  1.00&~~53  \\
 ~3.08 -- ~~4.61 &   3.77 &  12 & 2.56$\times10^7$    & 22.31$\pm$~6.44   &  1.00&~~36  \\
 ~4.61 -- ~~6.92 &   5.65 &   6 & 8.54$\times10^6$    & 20.49$\pm$~8.37   &  1.00&~~24  \\
 ~6.92 -- ~10.38 &   8.48 &   7 & 6.64$\times10^6$    & 43.92$\pm$16.60   &  1.00&~~18  \\
 10.38 -- ~15.57 &  12.71 &   4 & 2.53$\times10^6$    & 46.10$\pm$23.05   &  1.00&~~11 \\
 15.57 -- 119.23 &  42.91 &   7 & 2.24$\times10^5$    & 85.32$\pm$32.24   &  1.00&~~~7  \\

\hline

\end{tabular}
\end{table*}

\section{Survey Completeness and Source Counts}

The visibility area of the VIRMOS radio survey as a function of the peak
flux density S$_{\rm P}$
is shown in Fig.~\ref{area_limit}.  As expected, the visibility area
increases very rapidly between 0.05 and 0.09 mJy and becomes equal to
1 degree at S$_{\rm P}\geq$ 0.093 mJy. This is a consequence of the observing
strategy that has assured a very uniform noise over almost the entire 1 
square degree field used for the extraction of the catalogue.
Since the completeness of the radio sample is defined in terms
of the peak flux, while the source counts will be derived as
a function of the total integrated flux, corrections must be
applied to the observed numbers of radio sources in order to
take into account all possible observational biases. The most
important of such biases is probably the resolution bias
 which leads to missing
faint (i.e extended) sources at fluxes close to the limit. In 
fact, such sources, with peak flux densities below the
survey limit, but total integrated fluxes above this limit,
would not appear in the catalogue. The correction due to this
bias is a function of the intrinsic angular size distribution
of the sources and of the beam of the observations. To estimate
the correction factor to be applied to the observed data,
in the next section we
will model the effective angular-size distribution of the sources in
our radio sample and then we will use this distribution to simulate
a sample of radio sources that we will analyse with the same recipe
used  for the real sources described in Sect. 4.

\subsection{Angular Size Distribution}

Previous high-resolution studies of the faint radio population
suggested that the median angular size ($\theta_{med}$) for sub-mJy
radio sources is approximately 2$^{\prime\prime}$ and almost independent
of flux density between 0.08-1.0 mJy (\cite{Wind93}; \cite{Foma91};
\cite{Oort88}).
In order to derive an unbiased distribution of angular sizes
from our sample, we have to use sources in a range of total
fluxes in which the resolution bias is likely not to have modified,
in the catalogue, the intrinsic angular size distribution. For this
purpose we used all the sources with $0.4\le S < 1$ mJy. Given the
relation between angular size $\theta$ and the ratio between total and
peak fluxes, a source with $S_{\rm T}\simeq 0.6$ mJy would have a peak flux
greater than our detection limit even for a relatively large
angular size ($\theta\simeq 15$ arcsec). Forty-eight of the 111
sources (43\%) in this flux range are resolved, and we measured for
them a value of $\theta$. The other 63 sources are unresolved,
i.e. they are below the solid line representing
the upper envelope in Fig~\ref{ratio_fit_allflux}. For them we could derive
only an upper limit to their intrinsic size $\theta$; these upper limits
range from $\simeq 1.4$ to $\simeq 2.3$ arcsec. These $\theta$ values were then
analyzed with the survival analysis techniques of \cite{FN85},
using the statistical package ASURV (\cite{IFN86}).
This technique uses all the upper limits (which
are slightly more than $50\%$ in our data set of angular sizes)
in reconstructing the intrinsic distribution. The resulting
estimate for the median value of $\theta$ is $\theta_{\rm med} \simeq 1.8\pm
0.2$ arcsec, somewhat lower than, but consistent with the
value $\theta_{\rm med} \simeq 2.6\pm 0.4$ obtained by Richards (2000) for
sources in the same flux range from his deep survey in the Hubble Deep
Field region. Our value of $\theta_{\rm med}$ is also in good agreement
with the relation
$\theta_{\rm med} (^{\prime \prime}) =
2.0 S^{0.3}$, found by \cite{Wind90},
where S is the total flux in mJy.
In Fig.~\ref{theta_int_dist} we report the integral angular size
distribution for our sources with 0.4$\leq$S$<$1.0 mJy (solid line)
which, because of the presence of the upper limits, is determined only
for $\theta\ge 1^{\prime\prime}.8$.
The dashed line shows an analytical fit to this distribution
(h($\theta$)=1/(1.6$^{\theta}$) for $\theta\leq4^{\prime \prime}$
and h($\theta$)= ($\theta^{-1.3} - 0.01$) for $\theta > 4^{\prime \prime}$).
For comparison the dot-dashed line shows the integral angular
distribution h($\theta$)=exp[$-\ln 2 (\theta$/$\theta_{\rm med})^{0.62} $]
with $\theta_{\rm med}$ = 1.8 suggested by \cite{Wind90}.

As clearly shown in Fig.~\ref{theta_int_dist}, the Windhorst et al. relation
is not a good representation of our measured distribution of
angular sizes, because it predicts a substantial tail
of sources with large angular sizes, which is not present in
our data. For example, the fraction of sources with
$\theta > 4$ arcsec predicted by this relation
($\simeq 32\%$) is about twice as high as that measured from our data,
as well as that which can be derived for sources in the same
flux range from the radio data in the HDF region (\cite{Rich00};
see his Fig. 4). The use of the Windhorst relation, because
of its high fraction of sources with large angular sizes, would
lead to an overstimate of the correction factors due the resolution
bias. 
\begin{figure*}
\includegraphics[width=12cm]{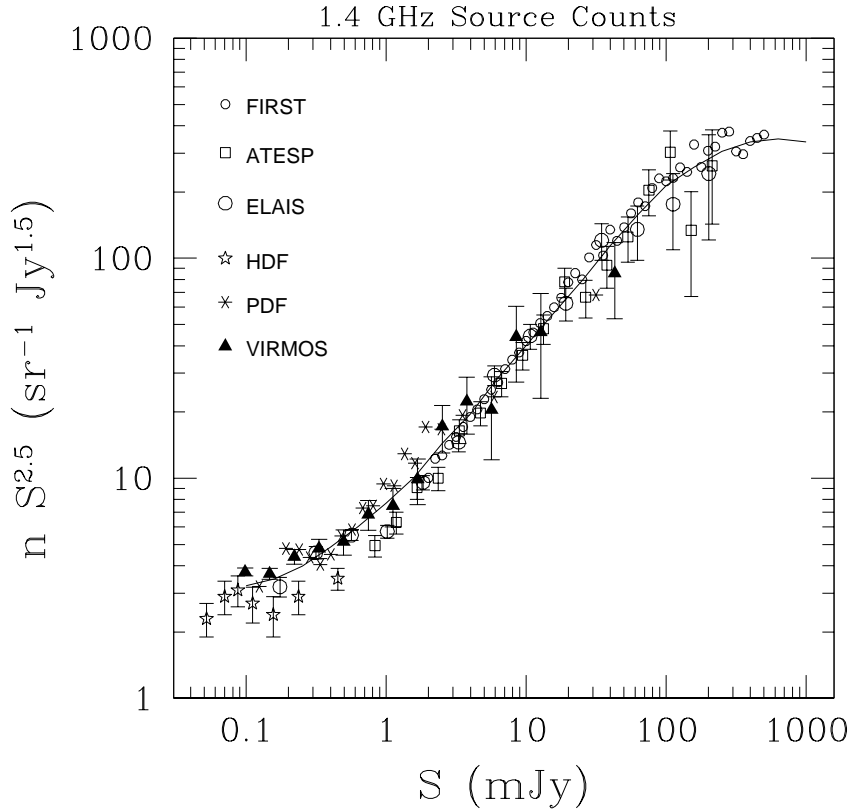}
 \centering
 \caption[] {The normalised differential source counts. The abscissa gives the
 flux density (mJy) and the ordinate gives the differential number
 of sources normalised by $S^{2.5}$ (sr$^{-1}$ Jy$^{+1.5}$).
 With different symbols are reported source counts at 1.4 GHz from several 
 surveys: the smaller open circles represent the counts from the FIRST survey
 (\cite{Whit97}), the open squares are the counts from the ATESP survey
(\cite{Pran01a}), the larger open circles are
 the counts obtained by the survey ELAIS (\cite{Grup99b};\cite{Cili99}),
the stars are the counts from the HDF (\cite{Rich00}), the asterisks are the 
counts from the Phoenix Deep Survey (\cite{Hopk98}), the
filled triangles are the counts obtained with the VIRMOS survey discussed
in this paper. The solid lines is a best fit to a compilation of earlier deep
surveys made by \cite{KOW88}
}
\label{source_counts}
\end{figure*}

\subsection{Completeness}

In order to estimate the combined effects of noise, source extraction
 and flux determination techniques and resolution bias on the completeness of
 our sample, we constructed simulated samples of radio sources down to a
 flux level of 0.04 mJy, i.e. a factor of two lower than the minimum
 flux we used to derive the source counts (see Sect. 5.3).
 This allows to take into account also those sources with an
 intrinsic flux below the detection limit which, because of
 positive noise fluctuations, might have a measured flux above
 the limit. The simulated samples have been extracted from
 source counts with the integral size distribution derived in the previous
section and described by a broken power law consistent with that observed.

Following these recipes, we simulated 9 samples, each of them
with a number of sources above the detection limit similar to
that observed in the real data (i.e. $\simeq 1000$). All the sources,
including those below the detection limit, were randomly injected
in the CLEANed sky images of the field and were
recovered from the image and their fluxes were measured using
the same procedures adopted for the real sources
(see Sect. 4). The detected simulated sources were then
binned in flux intervals. Finally, from
the comparison between the number of simulated sources detected in each
bin and the number of sources in the input simulated sample in the same flux
bin we computed the correction factor $C$ to be applied to our observed source
counts.  In 
Table~\ref{counts_tab} we report the average correction factor $C$ for each
flux density bin.
As expected, the resolution bias significantly affects the first two flux
density bins in the source counts. Our simulations tell us we are missing
about 29\%
and 25\% of sources in the first two flux density bins respectively.
In the bins at higher flux the results of these simulations are consistent
with no need for a correction. We therefore set $C=1.0$ for all these bins.

\subsection{Source counts}

In order to reduce possible problems near the flux limit and to avoid
somewhat uncertain corrections in the steep part of the visibility area
we constructed the radio sources counts considering only the 1013 sources with
a flux density greater than 0.08 mJy ($i.e.$ we excluded the 41 sources
with S$<$0.08 mJy).
The 1.4 GHz source counts of our
survey are summarised in  Table~\ref{counts_tab}.
For each flux density bin, the
average flux density, the observed number of
sources, the differential source density
(in sr$^{-1}$ Jy$^{-1}$), the normalised differential counts $nS^{2.5}$
(in sr$^{-1}$ Jy$^{1.5}$) with estimated Poissonian errors (as
$n^{1/2}S^{2.5}$).
In the last two columns we report the correction factor $C$ to be
applied to our source counts to correct for incompleteness and the
corrected integral counts (in deg$^{-2}$).

The normalised 1.4 GHz counts (column V) multiplied by the
correction factor $C$  are plotted in
Fig.~\ref{source_counts} where, for comparison, the
differential source counts obtained with other 1.4 GHz radio surveys
are also plotted. 

Our counts are in good agreement,
over the entire flux range sampled by our data (0.08 - 10 mJy),
with the best fit to earlier surveys (\cite{KOW88}).
It is interesting to note that,
because of the relatively good statistics of our data points
over about two orders of magnitude in flux, our data clearly
show the change in slope of the differential counts, occurring
below 1 mJy. Fitting the VLA-VDF differential and integral counts with two 
power laws we obtain, for $S<0.6$ mJy:
$$
{dn\over ds} = (57.54 \pm 1.07) \times S^{-2.28 \pm 0.04}
$$
and at higher fluxes, $S>0.6$ mJy:
$$
{dn\over ds} = (75.86 \pm 1.08) \times  S^{-1.79 \pm 0.05}
$$

where dn/ds is in deg$^{-1}$mJy$^{-1}$ e $S$ in mJy.
 At faint fluxes ($S < 0.4$ mJy) the most relevant comparison is
   with the results obtained by Richards (2000) in his deep VLA
   survey of a region of about 40 x 40 arcmin around the Hubble
   Deep Field. In this flux range the statistics in our data (i.e.
   number of radio sources) is about four times higher than that
   in the HDF data and our derived counts are about 50\% higher
   than those of Richards (see also \cite{Hopk02} for similar results). 
   This is consistent with the fact, already noted by Richards,
   that the counts in the HDF region appear to be sistematically
   lower than those of other fields above 0.1 mJy. This difference
   is probably due to a combination of real field-to field
   variations and some possible incompleteness of somewhat
   extended sources in the high resolution HDF data ($2^{\prime\prime}$ beam).
   Indeed, lower resolution, deep Westerbork observations of a
   sub-area in the Hubble Deep Field region have detected a
   non-negligible number of sources which do not appear in
   Richards' catalogue (\cite{Garr00}).

\section{Conclusions}

Using the VLA at 1.4 GHz we have observed a 1 deg$^2$ field centered
on the VIRMOS Deep Field (RA=02:26:00 DEC=$-04$:30:00),
imaging the whole area with uniform sensitivity and a
resolution of 6 arcsec.
We have investigated the effects of clean bias and bandwidth smearing
on our observations confirming that the observing
strategy and the data reduction procedure allow us to consider these effects
negligible.
A complete catalogue of 1054 radio sources down to a local
$5\sigma$ limit ($\simeq 80$ $\mu$Jy) has been compiled.
In order to assess the effects of random noise, source extraction technique
and resolution bias on the completeness of our sample we have first derived
the effective angular-size distribution using the sources in our sample in the
range 0.4-1.0 mJy. Then we have generated a large sample of simulated
sources using the derived angular-size distribution and extracted from
source counts described by a broken power law consistent with that observed.
These simulations allowed us to statistically correct ourc counts in
the faintest flux bins.
The final counts are in good agreement with the best fit to earlier surveys
(\cite{KOW88}). In particular, our data clearly show a significant change in
slope of the differential counts occurring below 1.0 mJy.
The best fit slope in the range 0.08-0.6 mJy ($-2.28\pm 0.04$) is close to
the Euclidean value. At faint fluxes ($S<0.4$ mJy), where we have a high
statistics, our derived counts are about 50\% higher than those of
\cite{Rich00} in the HDF region. This is consistent with the fact, already
noted by Richards, that the counts in the HDF region appear to be
sistematically lower than those of other fields above 0.1 mJy

The same region of the sky has been target of deep ($I_{\rm AB}=25$), 
multicolor (UBVRIK) photometry
observations. The photometric identification of most of the radio sources 
in the catalogue
and planned spectroscopic observations during the VIRMOS Deep Field Survey
will provide a unique opportunity to study the nature and properties of
the $\mu$Jy source population.

\begin{acknowledgements}
This work was performed under the framework of the VIRMOS consortium,
and was supported by the Italian Ministry for University and Research
(MURST) under grant COFIN-2000-02-34.

\end{acknowledgements}

\end{document}